\documentstyle[11pt,epsf,rotate]{article}
\newcommand{\bce}{\begin{center}}
\newcommand{\ece}{\end{center}}
\newcommand{\be}{\begin{equation}}
\newcommand{\ee}{\end{equation}}
\newcommand{\bea}{\begin{eqnarray}}
\newcommand{\eea}{\end{eqnarray}}
\newcommand{\bit}{\begin{itemize}}
\newcommand{\eit}{\end{itemize}}
\newcommand{\bfl}{\begin{flushright}}
\newcommand{\efl}{\end{flushright}}
\newcommand{\E}{\>=\>}
\newcommand{\EA}{&=&}
\newcommand{\To}{\> \longrightarrow \> }
\newcommand{\non}{\nonumber \\}

\renewcommand{\thesection}{\arabic{section}}

\begin{document}
\thispagestyle{empty}

\begin{flushright}
%PSI-PR-04-07
\end{flushright}

\vspace{3cm}

\bce
{\large \bf An Abraham-Lorentz-like Equation for the Electron from the}

\vspace{0.2cm}

{\large \bf Worldline Variational Approach to QED} 

\vspace{2cm}

\vspace{1.2cm}
R. Rosenfelder$ ^1$ and A. W. Schreiber$ ^{2,3}$

\vspace{0.8cm}
$^1$ Paul Scherrer Institute, CH-5232 Villigen PSI, Switzerland

\vspace{0.1cm}
\noindent
$^2$ Department of Physics and Mathematical Physics, and
           Research Centre for the Subatomic Structure of Matter,
           University of Adelaide, Adelaide, S. A. 5005, Australia

\vspace{0.1cm}

$^3$ Present address: Australian Centre for Plant Functional Genomics,
Hartley Grove, PMB 1 Waite Campus, University of Adelaide, Glen Osmond 5064
\ece

\vspace{2.5cm}

\begin{abstract}
\noindent
The variational equation for the mean square displacement of the 
electron in the polaron worldline approach to quenched QED can be cast 
into a form
which closely resembles the classical Abraham-Lorentz equation but without
the conceptual and practical diseases of the latter. The connection with
delay equations describing field retardation effects is also
established. As applications we solve this integro-differential equation
numerically for various values of the coupling constant and cut-off and  
re-derive the variational approximation to 
the anomalous mass dimension of the electron found recently. 

\end{abstract}

\newpage
\section{Introduction}

\setcounter{equation}{0}

There is a long history of attempts at classical models of the
electron until Quantum Electrodynamics (QED) took over as the most
successful and precise theory in the microscopic domain
\cite{Schweb}. Noteworthy among these models is, in particular, the
one due to Abraham and Lorentz nearly one hundred years ago which is
covered in many textbooks (see for example, Ref. \cite{Jack}). It describes 
a classical electron under the influence of both
an external  force as well as the back reaction from the energy loss 
due to radiation of photons, and leads to a 
third-order differential equation which,
however, has extraneous unphysical {\it run-away} solutions. The usual
 method to avoid these solutions is to convert the original equation 
into a second-order integro-differential equation.  The classic
nonrelativistic result is of the well-known form~\cite{Jack}
\be
\frac{d^2}{dt^2} \, {\bf x}(t) \E \frac{1}{m} \int_0^\infty ds \> 
e^{-s} \, {\bf F}(t+ s \tau)\;\;\;,
\label{eq: ab lorentz}
\ee
where $\tau$ is the characteristic time $2 e^2/3 m c^3$, $m$ is the
electron mass and ${\bf F}$ is
an external force which can depend on time either explicitly or implicitly
through ${\bf x}(t)$.  The solutions to this equation do not exhibit 
run-away behaviour, however Eq.~(\ref{eq: ab lorentz}) clearly violates 
(nonrelativistic) causality because the motion at time $t$ depends 
on forces at time $t'>t$.
It is noteworthy that the analogous equation, without runaway solutions,
for a relativistic Dirac particle \cite{Dirac} in 3 dimensions has only 
been derived rather recently \cite{IbPu}. 

\vspace{0.2cm}
From a modern perspective, of course, it is appropriate 
to wonder how nonrelativistic and classical equations like the one 
above are connected to a quantum mechanical and ultimately a quantum field
theoretical description.  In nonrelativistic QED this correspondence
limit has been examined in particular by Moniz and Sharp~\cite{MoSh}
while Johnson and Hu have derived an Abraham-Lorentz-Dirac equation as
a semi-classical limit within scalar quantum field theory~\cite{JoHu}.

In the present paper we discuss the fully relativistic field
theoretic problem of a spin-$1/2$ electron dressed by an arbitrary
number of photons in the quenched approximation to QED.  Our approach
is patterned after Feynman's celebrated variational treatment of the
polaron problem \cite{Feyn}, which was first applied by Mano
\cite{Mano} to a relativistic scalar field theory and re-discovered
and expanded by us in a series of papers \cite{WCiso,WC7}. Its main
features are the description of relativistic particles by worldlines
\cite{worldline} parametrized by the proper time, an exact functional
integration over the light fields (i.e. photons) and a variational
approximation of the resulting effective action by a retarded
quadratic trial action. In recent work we have extended this approach
to more realistic theories, in particular to quenched QED
\cite{QED1,QED2}.

The approach is fully nonperturbative, Lorentz covariant, respects
gauge symmetry and contains the exact one-loop self-energy in the
small coupling limit.  It does not rely on a perturbative, semi-classical 
or  a derivative expansion.  In Ref.~\cite{QED2} we concentrated on the
divergence structure and renormalization of the theory, resulting in a
compact expression for the anomalous mass dimension of the electron.
In the present work we go beyond this and calculate also the finite 
contributions. We do this by numerically solving the
variational equation for the position of the pole of the electron
propagator for a range of coupling constants and UV cut-offs.
  We shall also show that the relevant
variational equation  can be written as an
Abraham-Lorentz-like equation for the {\it mean square} displacement 
of the electron. As this ``Variational Abraham-Lorentz Equation'' (VALE)
turns out to be (at most) a second-order integro-differential
equation, no run-away solutions occur.  Indeed, anticipating the result
derived in the Section~\ref{sec: vale}, in general this equation
may be written in the suggestive form
\be
\frac{d^2}{dt^2} \left < x^2(t) \right > \E
2\, \left < \dot x^2(t) \right > \> + \> 
\frac{2}{\kappa_E} \int_{-\infty}^{+\infty} ds \> 
f(s) \left \{ \left <  \left[\Delta x(t)\right ]^2 \right>
\> - \> \left <  \left[\Delta x(t-s)\right ]^2 \right>\right \}\;.
\label{eq: var ab lorentz}
\ee
Here $x_\mu(t)$ is the Euclidean position at proper 
time $t$ ($t > 0$), $\Delta x^\mu(t)$ is its deviation from a classical
straight-line path, $\kappa_E$ is a reparameterization parameter 
playing the role 
of a mass and $<\ldots>$ refers to a certain averaging 
over all worldlines occurring in
the functional integral for the theory. These definitions will be made 
more precise in the following two Sections. At this stage we merely 
point out the marked similarity of this equation to 
Eq.~(\ref{eq: ab lorentz}), in particular if one re-writes the latter as
\be
\frac{d^2}{dt^2} \, {\bf x}^2(t) \E 2 \,\dot{\bf x}^2(t) \> + \> 
\frac{2}{m} \, {\bf x}(t) \cdot \int_0^\infty ds \> e^{-s}
{\bf F}(t+ s \tau)\;\;\;.
\label{eq: ab lorentz 2}
\ee

Of course, Eqs.~(\ref{eq: var ab lorentz}) and~(\ref{eq: ab lorentz 2}) 
are not completely identical, nor should they be.  In the absence of 
external forces the latter equation yields a differential equation 
with the solution that the particle moves uniformly along a straight 
line.  On the other hand, there \emph {are no} external forces in 
Eq.~(\ref{eq: var ab lorentz}).  The first term on the r.h.s. is the 
classical term while the integral characterizes an
`internal force' which has its origin in the constant emission and 
re-absorption of virtual photons.  This term is sensitive only to the 
(random-walk like) deviations from a straight line.

In the next Section we will describe the derivation of the variational
equations, leading to the VALE (\ref{eq: var ab lorentz}) in
Section~\ref{sec: vale} and will solve it numerically in
Section~\ref{sec: results}. Understanding these numerical results will 
also give a new,
simpler, derivation of the variational approximation to the anomalous
mass dimension than the one presented in Ref.~\cite{QED2}. 
Approximate forms of the VALE are discussed in an Appendix.

\section{The variational method}

In order to introduce our notation and terminology, we begin with a
brief summary of the essential points of the variational technique
applied to quenched QED and refer the reader to Refs.~\cite{WCiso,
QED1} and in particular Ref.~\cite{QED2} for the details. 

\subsection{Worldline formulation}

The starting point is the worldline, rather than field theoretic,
description for the propagator of a spinning particle which involves
functional integrals over a bosonic worldline $x_\mu(t)$ as well as a
Grassmannian $\zeta_\mu(t)$ which characterizes the spin.  
In Minkowski space time (with metric $(+,- - -$)) the free part 
(modulo boundary terms) reads
\be
S_0 \E \int_0^T dt \> \left [ \, - \frac{\kappa_0}{2} \, \dot x^2 + 
i \, \zeta \cdot \dot \zeta + \frac{1}{T} \, \dot x \cdot \zeta \chi \, 
\right ] 
\label{S0}
\ee
where $\kappa_0$ is the reparametrization parameter and 
$\chi$ the supersymmetric (SUSY) counterpart to the proper time $T$. 
The dependence of the action on these two degrees of freedom is connected
by a supersymmetric transformation \cite{QED1}. 
Due to time-translational invariance only the time interval $T$
matters and therefore in reality the integration limits 
in Eq. (\ref{S0}) and all following expressions are $ [t_0,  T+t_0] $ 
with $t_0$ arbitrary (e.g. $0$ or $-T/2$).
The Gaussian functional integral over the photon field in the 
interaction part can be carried out analytically, resulting in a 
bi-local effective action 
\be
S_1 \E - \frac{e^2}{2} \int_0^T dt_1 dt_2 \, 
\int \frac{d^4 k}{(2 \pi)^4} \, 
G^{\mu \nu}(k) \, J_{\mu}(k,x_1,\zeta_1) \,  J_{\nu}(-k,x_2,\zeta_2) \, 
e^{- i k \cdot (x_1 - x_2)}
\label{S1}
\ee
where $G^{\mu \nu}(k) $ is the photon propagator in an arbitrary 
(covariant) gauge,
\be
J_{\mu}(k,x,\zeta) \E \dot x_{\mu} - \frac{2}{\kappa_0} \zeta_{\mu} \, k 
\cdot \zeta
\label{current}
\ee
is the (convection and spin) current of the electron and 
$x_1 \equiv x(t_1) $ etc.
There is an elegant, compact and manifestly supersymmetric formulation 
in terms of ``superpositions'' and ``superderivatives'' \cite{QED1}
but in the following we will use the more cumbersome but explicit 
decomposition into bosonic and fermionic worldlines.

\subsection{Feynman-Jensen stationarity and variational principle}

Having integrated out the photons, the functional integrals over $x$ and 
$\zeta$ cannot be performed exactly
but may be approximated via the introduction of a trial action 
$S_t[x,\zeta]$ and by making use of Feynman's variational technique:
\be
\int e^{i S} \E \int e^{i S_t} \, \cdot \, 
\frac{\int e^{i S}}{\int e^{i S_t}} \E \int e^{i S_t} \, \cdot \,
\frac{\int e^{i (S-S_t)} e^{iS_t}} {\int e^{iS_t}} \, \equiv \,
\int e^{i S_t} \, \cdot \,
\left < e^{i (S-S_t)} \right >_{S_t} \> \simeq \> 
\int e^{i S_t} \, \cdot \, e^{i <S-S_t>_{S_t}} \> .
\label{eq: Jensen}
\ee 
Even with Euclidean times, where $ \> \exp(i S) \to \exp(-S_E) \> $, 
the usual Jensen inequality does not hold anymore since
we are also integrating over Grassmann-valued trajectories. However, only 
stationarity of the above expression under variations of the trial 
action is required in the following.
The subscript $S_t$ on the average reminds us that the weight
function is $e^{i S_t}$; the averages in Equation~(\ref{eq: var ab
lorentz}) are to be understood in the same way.  With an arbitrary
trial action the above approximation becomes exact at the stationary
point.  In practise, however, only trial actions at most quadratic in
$x$ and $\zeta$ can be used.  We work with a 
trial action of this form obtained by making the modified free action 
$ \tilde S_0 = S_0 + p \cdot x $ bi-local \footnote{The integration 
over the endpoint $x(T) = x$ of the trajectory with 
weight $\exp (i p \cdot x) $ (where $p$ 
is the external momentum of the particle) is included 
in the average. Note that the present retardation 
functions have a different normalization than in  Ref. \cite{WC7}. }
\be
\tilde S_t \E \lambda_1  \, p \cdot x + \int_0^T dt_1 \, dt_2 \> 
\left [ \, - \frac{\kappa_0}{2} \, g_B(\sigma) \, \dot x_1 \cdot 
\dot x_2 + i \, g_F'(\sigma) \, \zeta_1 \cdot \zeta_2 - 
\frac{1}{T} \, \sigma \, g_M'(\sigma) \, \dot x_1 \cdot \zeta_2 
\chi \, \right ] 
\label{St}
\ee
In addition to a scalar variational parameter $\lambda_1$ 
this trial action contains three
arbitrary (but even) variational `retardation' functions of 
$\sigma = t_1-t_2$ multiplying the three quadratic combinations 
$\dot x \,\dot x$ (i.e. `bosonic', or $B$), $\zeta\,\zeta$ 
(`fermionic', $F$) and $\dot x \, \zeta$ (`mixed', $M$).
Only the first two of these are relevant in the calculation of the
pole position of the electron propagator, which is obtained for 
$ T \to \infty$. An explicitly supersymmetric trial action would 
require only one retardation function 
\be
g_B(\sigma) \E  g_F(\sigma) \E g_M(\sigma) \> \equiv \> g(\sigma) \;\;\;.
\label{susy rel for ret}
\ee
As the exact action contains 
SUSY-breaking boundary terms (which will be discussed in more detail 
elsewhere \cite{RS})
we shall not enforce Eq.~(\ref{susy rel for ret}). 
It is therefore advisable to
take the most general {\it ansatz} 
and let the variational principle
choose the optimal solution within the given class of test functions
\footnote{It should be noted that Eq. (\ref{St}) is still not the most 
general quadratic ansatz as the retardation functions are assumed 
to depend only on the proper time difference and additional 
Lorentz structures are absent. For the general case in a scalar theory 
see Ref. \cite{WC7}, Appendix C.}.
Note that for $\lambda_1 = 1, g_B (\sigma) = g_F(\sigma) = g_M(\sigma) = 
\delta(\sigma) $ the trial action (\ref{St}) reduces to the 
free action (\ref{S0}). Because averaging with the free action is 
equivalent to first-order perturbation theory this implies that the 
variational approach gives the correct one-loop self energy for 
small coupling.

\subsection{Mano's equation}

The averages involved in the last term of Eq.~(\ref{eq: Jensen}) may be 
separated into averages of $S_0 - S_t$ (which, together with 
$ \int \exp (iS_t) $, may be combined into a quantity $\Omega$) 
and the average of $S_1$ (denoted by $V$). 
The roles of $\Omega$ and $V$ are not
unlike those of, respectively, kinetic and potential contributions in 
a standard quantum mechanical variational calculation.
Near the pole the electron propagator takes the form
\be
G_2(p) \To Z_2 \, \frac{ p \hspace{-5pt}/ + M}{p^2 - M^2}
\label{G2 pole}
\ee
and the on-shell limit of the argument of the exponential of the last 
term in Eq.~(\ref{eq: Jensen}) 
directly yields the relationship between
the electron's bare mass  $M_0$ and its physical mass $M$.  Explicitly, 
this equation, termed `Mano's equation', becomes
\be
M_0^2 \E M^2 ( 2 \lambda - \lambda^2) - 2 \Biggl ( \> \Omega[A_B] 
- \Omega[A_F]  + V [\mu^2_B,\mu^2_F]  \> \Biggr ) \;\;\;.
\label{Mano eq}
\ee
Here the `profile functions' $A_B$ and $A_F$ are Fourier transforms
of the bosonic and fermionic variational retardation functions, 
respectively:
\be
A_i(E) \E \int_{-\infty}^{+\infty} d\sigma \> g_i(\sigma) \, 
e^{i E \sigma} \E 
2 \, \int_0^{\infty} d\sigma \> g_i(\sigma) \, \cos \left ( E \sigma 
\right ) \> ,  \hspace{2cm}i \E B,\> F\> \> . 
\ee
They are fixed, as is $\lambda \equiv \lambda_1/A_B(0) $, 
through the Feynman-Jensen variational principle 
which  guarantees that Mano's equation (\ref{Mano eq}) is stationary 
w.r.t. their variation.   Once fixed, all quantities of interest in the 
field theory (e.g. masses, form factors, scattering cross 
sections~\cite{WCiso}) 
may be expressed in terms of these.  $A_B$ and $A_F$ only appear 
implicitly in the potential $V$,  through the bosonic and fermionic 
`pseudotimes' $\mu^2_{B,F} (\sigma)$ defined by
\be
\mu^2_i (\sigma) \E \frac{4}{\pi} \int_0^{\infty} dE \> 
\frac{1}{E^2 A_i(E)} \, \sin^2 \left ( \frac{E \sigma}{2} \right ) .
\label{def pseudotime}
\ee

\subsection{Kinetic and potential terms}
\label{subsection: kin pot}

Explicitly, the kinetic terms $\Omega[A_i]$ are given, in 4-dimensional
Euclidean space, by
\be
\Omega[A_i]\E \frac{2 \kappa_E}{\pi} \int_0^\infty dE
\left (\log {A_i(E)} \> + \> \frac{1}{A_i(E)}\>-\>1\right )\;\;\;.
\ee
Here $\kappa_E > 0$ is a parameter which re-parameterizes the
proper time without changing the physics.  It is useful to keep it because
it plays the role of a mass in the worldline description (see Footnote~3 
in Ref.~\cite{WC7}; the Euclidean parameter $\kappa_E$ is related to its
Minkowski counterpart $\kappa_0$ through $\kappa_0 = i\kappa_E$). 

The potential term $V$ results from averaging the bi-local effective 
action (\ref{S1}) and performing the 
limit $ T \to \infty$. While details of this
calculations are rather involved and will be given elsewhere \cite{RS} 
the final result was already presented in  Ref.~\cite{QED2}.
As $V$ is ultraviolet (UV) divergent we write it down in
$d = 4-2\epsilon$ dimensions.  
In Euclidean time (with $ p_E^2 = - M^2 $ ) it takes the form
\bea
V \EA V_1 + V_2 
\label{V=V1+V2}\\
V_1 \EA - \frac{ \pi \alpha}{\kappa_E} \, \nu^{2 \epsilon} \, (d - 1) \, 
\int_0^{\infty} d\sigma \> \int \frac{d^dk}{(2 \pi)^d} \> 
\frac{k^2}{k^2 + m^2} \, \left [ \, \left ( 
\dot \mu^2_F(\sigma) \right)^2 - \left ( \dot \mu^2_B(\sigma) 
\right)^2 \, \right ]  \, E(\sigma,k) 
\label{V1 by k,sigma} \\
V_2 \EA  \frac{ 4 \pi \alpha}{\kappa_E} \, \nu^{2 \epsilon} \, \lambda^2 
\int_0^{\infty} \! d\sigma \int \frac{d^dk}{(2 \pi)^d} \,
\frac{1}{k^2 + m^2} \, \left [ \, M^2   
- (d-2) \, \frac{ (k \cdot p_E)^2}{k^2} \, \right ] \, E(\sigma,k)  \> .  
\label{V2 by k,sigma} 
\eea
Here $\nu$ is an arbitrary mass parameter, the abbreviation
\be
E(k,\sigma) \E \exp \left \{ - \frac{1}{2 \kappa_E} \left [ k^2 
\mu^2_B(\sigma) - 2 \lambda k \cdot p_E \,\sigma \right ] \right \}
\ee
has been used and a photon mass $m$ has been kept in the photon 
propagator. The separation of $V$ into two parts makes sense because 
of several facts: first, it is seen that $V_2$ is more singular than 
$V_1$, having an additional power $k^2$ in the integrand. Second, 
$V_2$ vanishes for massless electrons $M = 0$, is repulsive 
($V_2 > 0$) and only depends on the bosonic pseudotime.

Previously the anomalous mass dimension $\gamma_M$ was calculated 
analytically in dimensional regularization~\cite{QED2}.  However, from 
a numerical point of view this regularization is extremely cumbersome 
in nonperturbative calculations (see, for example, the Dyson-Schwinger 
equation studies in Ref.~\cite{DS1}) and is usually replaced by a 
simple momentum cut-off.  Proper time regularization, i.e. a lower 
cut-off of proper time integrals at 
$\sigma = 1/\Lambda^2$, $\Lambda \rightarrow \infty$
could also be used (it would maintain translational invariance), but it
violates reparametrization invariance.
In the present work, it is much more convenient to make use of a 
form factor which is very similar to the  nonlocal regularization method 
proposed in Ref. \cite{EMKW}. 
Here the main reason is that the UV divergence in $V$ originates 
in an undamped (for $\sigma\rightarrow 0$, in which case 
$\mu^2_i(\sigma) \rightarrow \sigma$) 
Euclidean momentum integral over the exponential factor $E(k,\sigma)$.
This is the only appearance of $\mu^2_B$ in $V$ and hence,
after introducing a form factor of the form
\be
F(k^2) \E \exp \left (- \frac{k^2}{2 \Lambda^2} \right ) \> , 
\label{form factor}
\ee
all dimensionally regulated expressions in Ref.~\cite{QED2} may be 
converted to form factor regulated ones by setting $d = 4$ and making 
the simple replacement
\be
\mu^2_B(\sigma) \To \tilde \mu^2_B(\sigma) \E \mu^2_B(\sigma) + 
\frac{\kappa_E}{\Lambda^2}\;\;\;.
\label{def mu2 tilde}
\ee
With this regularization understood, the different terms in the
potential $V$ of Eq. (\ref{V=V1+V2})
become (after performing the momentum integration for $m = 0$)
\bea
V_1[\mu^2_B,\mu^2_F] \EA - \frac{3 \alpha}{4 \pi} \, \kappa_E 
\int_0^{\infty} d\sigma \, \frac{ (\dot \mu_F^2(\sigma))^2 - 
(\dot \mu_B^2(\sigma))^2}{\tilde \mu_B^4(\sigma)} \, 
e^{- \tilde \gamma(\sigma)} 
\label{V1 form}\\
V_2[\mu^2_B] \EA \frac{3 \alpha}{\pi} \kappa_E \int_0^{\infty} 
\frac{d\sigma}{\sigma^2} \, 
\frac{1-[1+ \tilde \gamma(\sigma)] e^{- \tilde \gamma(\sigma)}}
{\tilde \gamma(\sigma)} \> \> .
\label{V2 form}
\eea
Here 
\be
 \tilde \gamma(\sigma) = \frac{\lambda^2 M^2 \sigma^2}
{2 \kappa_E \, \tilde \mu^2_B(\sigma)}\;\;\;,
\hspace{3cm}
\dot \mu^2(\sigma) \equiv \frac{d \mu^2(\sigma)}{d\sigma}\;\;\;
\ee
and  $\alpha = e^2/(4 \pi) $ (in the real world $\simeq 1/137$ ) 
is the fine structure constant.

It is clear from Eqs.~(\ref{V1 form},\ref{V2 form}) that the 
$\sigma \to 0$ UV-divergences due to inverse powers 
of the bosonic pseudotime are now regulated and that both $V_1$ and $V_2$ 
as well as  $\Omega_i$ do
not depend on the reparametrization parameter $\kappa_E$, which only sets 
the scale
for the $\sigma$ and  $E$-variables \footnote{This is due to the 
reparametrization dependence 
of $ A_i(E,\kappa_E) = A_i(\kappa_E E)$ and $ \mu^2_i(\sigma,\kappa_E) = 
\kappa_E \> \mu^2_i(\sigma/\kappa_E)$ .} .
One also sees that bosonic and fermionic degrees of
freedom do not enter symmetrically in the interaction 
(e.g. $V_2$ only depends on $\tilde \mu^2_B$) and therefore the 
pseudotimes are in general different. 
This is due to supersymmetry violation by boundary terms
in the trial action and could affect the spin structure
of the propagator in our variational scheme. 
Since the spin-dependent terms are subleading in the limit 
$ T \to \infty$ and are not considered in the 
present investigation we just let the variational principle 
decide how much of SUSY violations it tolerates with the present 
trial action. Fortunately, as we will observe numerically 
and is to be expected from boundary terms, these supersymmetry 
violations are restricted to large values of $\sigma$ (or small values 
of $E$) and therefore do not have any influence on the UV behaviour of 
the solutions. 

Because the fermionic contributions, both in
the kinetic term of Mano's equation~(\ref{Mano eq})
as well as in $V_1$ (Eq.~(\ref{V1 form})), appear with
an opposite sign to the bosonic contributions they would cancel them
for exact supersymmetry, i.e.  $\mu_F^2(\sigma) = \mu_B^2(\sigma)$.
Note that for $\sigma \rightarrow 0$ a cancellation of this
sort is absolutely necessary as otherwise one 
would encounter the same quadratic divergences which occur in the 
propagator of scalar QED. In the context of the variational calculation 
the cancellation occurs only  {\it after} variation because one needs 
a restoring force in the variational principle. A simple example 
illustrating this subtlety is given in Ref. \cite{SRA workshop}.

\subsection{Variational equations}

With the explicit expressions for $\Omega$ and $V$ above it is now a
simple, albeit somewhat tedious, exercise to derive the variational
equations for $\lambda$, $A_F$ and $A_B$.  One obtains, respectively,

\be
\lambda \E 1 - \frac{3 \alpha}{2 \pi} \frac{\kappa_E}{\lambda M^2} \, 
\int_0^{\infty} d\sigma \> \Biggl [ \> 
\frac{ (\dot \mu_F^2)^2 - (\dot \mu_B^2)^2}{\tilde \mu_B^4} \, 
\tilde \gamma e^{-\tilde \gamma} + \frac{4}{\sigma^2} 
\frac{(1+\tilde \gamma+\tilde \gamma^2)e^{-\tilde \gamma} -1}
{\tilde \gamma}
 \> \Biggr ] \;\;\;,
\label{var eq lambda 1}
\ee
\be
1 - \frac{1}{A_i(E)} \E \frac{2}{ \kappa_E} \,  \int_0^{\infty} 
d\sigma \> \left ( -  \right)^i\frac{\delta V}{\delta \mu^2_i(\sigma)} \, 
\frac{\sin^2(E \sigma/2)}{E^2 A_i(E)} 
\label{rel a}
\ee
where we define $ (-)^B = 1$  and $(-)^F = - 1 \> $, the difference
in sign originating from the opposite sign of $\Omega[A_B]$ and
$\Omega[A_F]$ in Mano's equation.
The functional derivatives in Eq.~(\ref{rel a}) are
\be
\frac{\delta V}{\delta \mu^2_F(\sigma)} \E
\frac{3 \alpha}{2 \pi} \kappa_E 
\>\frac{d}{d\sigma}
\left [ \, \frac{\dot \mu_F^2(\sigma)}{\tilde \mu_B^4(\sigma)} 
e^{-\tilde \gamma(\sigma)} \, \right ]
\label{var eq for muF}
\ee
and
\bea
\frac{\delta V}{\delta \mu^2_B(\sigma)} \EA -  \frac{3 \alpha}{2 \pi} \kappa_E 
\, \frac{d}{d\sigma}
\left [ \, \frac{\dot \mu_B^2(\sigma)}{\tilde \mu_B^4(\sigma)} \>
e^{-\tilde \gamma(\sigma)} \, \right ] + \frac{3 \alpha}{4 \pi} \kappa_E \,
\frac{1}{\tilde \mu_B^2} \, \Biggl [ \> 
\frac{ (\dot \mu_F^2)^2 - (\dot \mu_B^2)^2}{\tilde \mu_B^4} \, 
(2-\tilde \gamma) e^{-\tilde \gamma} \non 
&& \hspace{7cm} - \frac{4}{\sigma^2} 
\frac{(1+\tilde \gamma+\tilde \gamma^2)e^{-\tilde \gamma} -1}
{\tilde \gamma} \> \Biggr ] \> .
\label{var eq for muB}
\eea
In the limit in which the interactions are turned off, we have $A_i(E) =
\lambda = 1$ and therefore $ \mu^2_i(\sigma) = \sigma $, $\Omega_i =
0 , M = M_0$. In the interacting theory the pseudotimes
$\mu_i^2 (\sigma)$ still have linear
behaviour for both $\sigma \to 0, \, \infty $. From 
Eqs. (\ref{var eq for muF},\ref{var eq for muB}) we then see that
without regularization
\be 
\frac{\delta V_1}{\delta \mu^2_i (\sigma)} \To 
\frac{\rm const.}{\sigma^3} \> , 
\hspace{1cm}  \frac{\delta V_2}{\delta \mu^2_B (\sigma)} \To 
\frac{\rm const.}{\sigma^2}  
\ee
for $\sigma \to 0 $.
It is this more singular UV-behaviour of $\delta V_1$ which requires 
regularization also in the variational equations. This is the trademark 
of a renormalizable theory whereas the more benign behaviour of 
$\delta V_2$ is characteristic for a super-renormalizable theory. 
On the other hand in the infrared region $V_2$ dominates because 
the very last term in Eq. (\ref{var eq for muB}) is not exponentially 
suppressed
\be
 \frac{\delta V_2}{\delta \mu^2_B (\sigma)} \> 
\stackrel{\sigma \to \infty}{\longrightarrow} \> 
\frac{\rm const}{\sigma^4} \> .
\label{V2 ir}
\ee
This is due to taking a vanishing photon mass $m = 0 $
in Eq. (\ref{V2 by k,sigma}).

\section{ The variational Abraham-Lorentz like equations}
\label{sec: vale}

It is obvious that the profile functions $A_i$ and the
pseudotimes $\mu^2_i$ contain the same information, as seen in
Eq.~(\ref{def pseudotime}). It would be more efficient, therefore, to
eliminate one of these from the variational equations (\ref{rel a}).
We shall now show that it is possible to rewrite Eq.~(\ref{rel a})
entirely in terms of the pseudotimes.

A differentiation of Eq. (\ref{def pseudotime}) with respect to $\sigma$
yields
\be
\dot \mu^2_i(\sigma) \E \frac{d}{d \sigma} \, |\sigma | + \frac{2}{\pi} 
\int_0^{\infty}
dE \> \frac{\sin (E \sigma)}{E} \left ( \frac{1}{A_i(E)} - 1\right ) \> .
\ee
It is important here to realize that the pseudotime is even (as can be 
seen from Eq. (\ref{def pseudotime})) and that 
the first term on the r.h.s. therefore gives sgn $(\sigma)$. The 
remaining integral is convergent due to subtraction of the asymptotic 
value $1/A_i(\infty) = 1$. Another differentiation leads to
\be
\ddot \mu^2_i(\sigma) \E 2 \, \delta(\sigma) +  \frac{2}{\pi} 
\int_0^{\infty} dE \> \cos(E \sigma) \left ( \frac{1}{A_i(E)} - 1 
\right ) \> .
\ee 
If we now insert Eq. (\ref{rel a}) we obtain
\be
\ddot \mu^2_i(\sigma) \E 2 \, \delta(\sigma) - \frac{2}{\pi} \, 
\frac{1}{\kappa_E} \int_0^{\infty} dE \, \int_0^{\infty} d\sigma' \> 
\left ( -  \right)^i\frac{\delta V}{\delta \mu^2_i(\sigma')}
\, \frac{1}{E^2 A_i(E)} \, 
\Bigl [ \, 1 - \cos(E \sigma') \, \Bigr ] \, \cos(E \sigma) \> .
\ee
By using the addition theorem for the cosine function  and the definition 
(\ref{def pseudotime}) the $E$-integration can be
performed exactly and gives
\be
\ddot \mu^2_i(\sigma) - \frac{1}{\kappa_E} \int_0^{\infty} d\sigma' \>
\left ( -  \right)^i \,
\frac{\delta V}{\delta \mu^2_i(\sigma')} \, \left [ \, \mu^2_i(\sigma) - 
\frac{1}{2} \mu^2_i\left ( \sigma + \sigma' \right ) - \frac{1}{2} 
\mu^2_i \left ( \left |\sigma - \sigma' \right | \right ) \, \right ] 
\E  2 \, \delta(\sigma) \> , \> \> \sigma \ge 0 \> \> .
\label{AL 1}
\ee
The appropriate boundary conditions for solutions to this 
integro-differential equation
are
\be
\mu^2_i(0) \E 0 \> ,\hspace{0.5cm} \lim_{\sigma \to +0} 
\dot \mu^2_i(\sigma) \E 1 \>.
\label{bound cond}
\ee
The proper times $\sigma, \sigma'$ are restricted to be nonnegative and 
we therefore have to take the absolute value in the argument of the 
shifted pseudotime in Eq.~(\ref{AL 1}). 
Alternatively, this restriction may be avoided by
remembering that $\mu^2_i(\sigma)$ is an even function of $\sigma$  
and hence the integrands appearing in the expression for $V$
(Eqs.~(\ref{V1 form},\ref{V2 form})) are also.  One  obtains
\be
\ddot \mu^2_i(\sigma) - \frac{1}{2 \kappa_E} \int_{-\infty}^{+\infty} 
d\sigma' \> \left ( -  \right)^i 
\frac{\delta V}{\delta \mu^2_i(\sigma')} \, \left [ \, \mu^2_i(\sigma) - 
\mu^2_i(\sigma - \sigma') \, \right ] \E  2 \, \delta(\sigma) \> , 
\> \> - \infty \le \sigma \le + \infty \> \> 
\label{AL 2}
\ee
and, in fact, even the $\delta$-function may be eliminated by performing
differentiation with respect to $|\sigma|$ rather than $\sigma$:
\be
\mu^{2 \, \prime \prime}_i(|\sigma|) - \frac{1}{2\kappa_E} 
\int_{-\infty}^{+\infty} 
d\sigma' \> \left ( -  \right)^i 
\frac{\delta V}{\delta \mu^2_i(|\sigma'|)} \, \left [ \, 
\mu^2_i(|\sigma|) - \mu^2_i(|\sigma - \sigma'|) \, \right ] \E  0 \> , 
\> \> - \infty \le \sigma \le + \infty \> \> 
\label{AL 3}
\ee
Here $''$ denotes differentiation with respect to the argument. 

We shall present numerical solutions to these variational equations 
(which, as promised, no longer involve the profile functions directly) 
in the next Section.  At this stage, however, we would first like to 
discuss the meaning of Eq.~(\ref{AL 3}) for the bosonic case, 
i.e. $i = B$.  In Ref.~\cite{WC7}, Eqs. (25,26) the expectation value of
$x^\mu(t_1)-x^\mu(t_2)$ and $[x^\mu(t_1)-x^\mu(t_2)]^2$, when 
averaged with the trial action, were calculated.  The corresponding 
results for the present case (i.e. QED, with Euclidean metric, and 
setting $t_2=0$ and hence $x^\mu(t_2)=0$ for convenience) become
\bea
\left <|x(t)|\right>_{S_t} \EA \frac{\lambda \,M}{\kappa_E}\> t 
\>\equiv\> \left|x_{\rm class.}(t) \right| \non
\left <x^2(t)\right>_{S_t} \EA \left <|x(t)|\right>_{S_t}^2 \> + \> 
\frac{4}{\kappa_E} \>\mu^2(t)\>\equiv\> 
x_{\rm class.}^2(t) \> + \> 
\left <\left [\Delta x(t)\right]^2\right>_{S_t}\;\;\;,
\label{eq: mu interpretation}
\eea
i.e.  the mean squared displacement of the electron is made up by 
an overall quadratic drift due to the electron's momentum, i.e. 
growing like $t^2$, and a term proportional to the pseudotime, 
growing like $t$ for small $t$;  the latter characterizes both 
the quantum mechanical Brownian motion as well as the continual 
random ``kicks'' from emission and absorption
of virtual photons in the cloud surrounding the bare particle. 

Inserting Eq.~(\ref{eq: mu interpretation}) into the variational equation
(\ref{AL 3}) results in the {\it Abraham-Lorentz}-like
equation mentioned in the Introduction, i.e.
\be
\frac{d^2}{d|t|^2} \left < x^2(t) \right >_{S_t} \E
2\, \left < \dot x^2(t) \right >_{S_t} \> + \> 
\frac{2}{\kappa_E} \int_{-\infty}^{\infty} dt' \> \frac{1}{4}
\frac{\delta V}{\delta \mu^2_i(t')}
\left \{ \left <  \left[\Delta x(t)\right]^2 \right>_{S_t}\> - \>
\left <  \left[\Delta x(t-t')\right]^2 \right>_{S_t}\right \}\;.
\label{eq: var ab lorentz 2}
\ee
$ \delta V/ \delta \mu^2_i $ may then be interpreted as 
(internal) ``force'' acting on the bare electron. 

\vspace{0.2cm}
As mentioned in the Introduction, the original Abraham-Lorentz
equation actually is a third-order linear differential equation for
the position of the electron rather than an integro-differential
equation. It is possible to write the VALEs in Eqs. 
(\ref{AL 1}) - (\ref{AL 3}) in a similar way by converting them 
into a special form of {\it delay} equations.
This class of differential equations has been 
studied extensively for problems of radiation damping in electrodynamics 
and general relativity and in mathematics \cite{delay}. 
Indeed, by invoking the first mean value theorem (this requires 
that $\delta V/\delta \mu^2_i(\sigma) $ is of one sign, which appears 
to be fulfilled) to the integral in Eq. (\ref{AL 3})
one can replace the $\sigma'$-argument of the pseudotime by some 
mean time $ 0 \le \tau_i \le \infty$. The remaining integral then 
just gives a constant 
\be
C_2^i  \E \int_0^{\infty} d\sigma \> ( - )^i \frac{\delta V}{\delta 
\mu^2_i(\sigma)} \>.
\ee
Therefore we have for $\sigma \ge 0$
\be
\ddot \mu^2_i(\sigma) - \, \frac{1}{\kappa_E} C_2^i \, \left [ \, 
\mu^2_i(\sigma) - 
\frac{1}{2} \mu^2_i\left ( \sigma + \tau_i \right ) - \frac{1}{2} 
\mu^2_i \left ( \left |\sigma - \tau_i \right | \right ) \, \right ] 
\E  0 \> .
\label{delay eq}
\ee
The above (seemingly linear) delay equation is exactly equivalent to 
the original (nonlinear) VALE because the constant $C_2^i$ is in general 
a functional of the pseudotime and because by construction the delay 
$\tau_i$ will also depend on the external proper time $\sigma$ as well 
as on $ \mu^2_i(\sigma) $. It is worth noticing that $C_2^i$ also 
governs the asymptotic behaviour of the profile function 
\be
A_i(E \to \infty) \E 1 + \frac{C_2^i}{(\kappa_E E)^2} + \ldots \> .
\ee
By differentiating Eq. (\ref{AL 3}) it is seen that it also
determines the initial ``jerk'' \cite{Leen}:
\be
\stackrel{\ldots}{\mu}^2_i(\sigma \to +0) \E  \frac{C_2^i}{\kappa_E} \> .
\ee 
For the fermionic case (or in the supersymmetric limit) 
$\delta V/\delta \mu^2_F$ 
is a total derivative (see  Eq.~(\ref{var eq for muF})), so here one 
finds the exact expression $C_2^F = 3 \alpha \Lambda^4/(2 \pi)$.

In many applications of delay equations a constant delay is
assumed. Although this does not seem to be a valid approximation for
QED it is clear that the delay $\tau_i$ is a remnant of the photon
degrees of freedom which have been integrated out and no longer appear
 in the equation of motion for the mean displacement of the
bare electron. Since in QED the ``force'' $ \delta V/\delta \mu^2_i $
is very singular at small $\sigma$ one  expects a very small delay
$\tau \sim 1/\Lambda^2$ (all integrals are dominated by small
values of $\sigma'$): the average time a virtual photon is ``in the air'' 
is very short, but the kick given to the bare electron is very violent.

Finally, we emphasize that in the discussion above no use has been made 
of the specific form of the interaction, i.e. the 
VALEs (\ref{AL 1}) - (\ref{AL 3}) 
hold for any dressed particle.  If we concentrate now on the specific 
form (\ref{V1 form},\ref{V2 form}) of the interaction in QED 
further simplifications can be made. For investigations of the
divergence structure of the theory as in  Ref.~\cite{QED2} it should be a
reasonable approximation to neglect $V_2$ since it it was shown to be 
less singular than $V_1$. Then
$\delta V/\delta \mu_i^2(\sigma)$ is just a total derivative 
(see Eq.~(\ref{var eq for muF}))
and one may integrate the VALE with respect to $\sigma$. 
In this supersymmetric approximation fermionic and bosonic pseudotimes
are identical and for $\sigma \ge 0 $ obey the equation 
\be
\dot \mu^2_{\rm SUSY}(\sigma) + \frac{3\alpha}{4\pi} \, 
\, \int_0^{\infty} d\sigma' \, \frac{\dot \mu^2_{\rm SUSY}(\sigma')}
{[\tilde \mu^2_{\rm SUSY}(\sigma')]^2}
\, e^{-\tilde \gamma(\sigma')} \, 
\left [ \, \mu^2_{\rm SUSY}(\sigma+\sigma') - 
\mu^2_{\rm SUSY}(|\sigma - \sigma'|) \, \right ] \E  1 \> .
\label{AL 4}
\ee
Since the regularized pseudotime just adds a constant to the pseudotime 
$\mu^2(\sigma)$, Eq. (\ref{AL 4}) also holds for $\tilde \mu^2(\sigma) $.
To exhibit the UV divergences one may even neglect the 
electron mass altogether. In this ``asymptotic SUSY'' (ASUSY) 
approximation the quantity $\tilde \gamma$ in Eq. (\ref{AL 4}) is set to 
zero which allows a further  integration over $\sigma$. 
Taking into account the boundary condition $\mu^2(0) = 0$ 
one obtains the simpler integral equation ($\sigma \ge 0$)
\be
\mu^2_{\rm ASUSY}(\sigma) + \frac{3 \alpha}{4 \pi} \, 
\, \int_0^{\infty} d\sigma' \>  \frac{\mu^2_{\rm ASUSY}(\sigma+\sigma') 
+  \mu^2_{\rm ASUSY}(|\sigma - \sigma'|) - 2  \, 
\mu^2_{\rm ASUSY}(\sigma')}
{\tilde \mu^2_{\rm ASUSY}(\sigma')} \E  \sigma \> .
\label{AL 5}
\ee
Adding $\kappa_E/\Lambda^2$
on both sides turns it into a compact, but still nonlinear integral 
equation for 
\be
y \left (s = \frac{\sigma \Lambda^2}{\kappa_E} \right ) \> := \>  
\frac{\Lambda^2}{\kappa_E} \,  \tilde \mu^2_{\rm ASUSY}(\sigma) 
\ee
namely,
\be
y(s) \E 1 + |s| - \frac{3 \alpha}{2 \pi} \, \int_{-\infty}^{+\infty} 
ds' \>  \left [ \, \frac{y(s + s')}{y(s')} - 1 \, \right ] \> . 
\ee
We have been unable to find an exact analytical solution for this 
scale-free equation. For the purpose of calculating the anomalous mass 
dimension it is, however, sufficient to know the asymptotic behaviour 
of $y(s)$ which will be derived below. An approximate,
third-order differential equation (i.e. of the same type as the original 
Abraham-Lorentz equation) is obtained in Appendix \ref{app: 3rd order de}.

\section{Numerical results}
\label{sec: results}

Without any approximation the three variational equations, i.e. the
fermionic and bosonic VALEs (Eq.~(\ref{AL 1})) together with the 
variational equation for $\lambda$
(Eq.~(\ref{var eq lambda 1})) may be solved iteratively by stepwise 
numerical integration over $\sigma$. Because the profile functions
$A_{F,B}$ have been eliminated in favour of the pseudotimes, it is not
necessary to numerically evaluate the cosine transform (Eq.~(\ref{def
pseudotime})) for the pseudotime and its derivative.  This simplifies
the numerical effort as compared to the published variational
treatments of super-renormalizable and finite theories
\cite{WCiso,WC7,polaron_aniso}. 

However, the presence of two vastly
different scales characterising both the ultraviolet (~$\sigma \sim
\kappa_E/\Lambda^2$~) and the infrared region (~$ \sigma \sim
\kappa_E/(\lambda M^2) $ with $\lambda \to 0$ for $\Lambda \to
\infty$, see below) makes the calculation much harder for the present
renormalizable theory.
Stable numerical results have been obtained for cut-offs as large as $
\Lambda = 500 \, M$ and coupling constants up to $\alpha = 0.5 - 1.0
$ and have been tabulated in Table~\ref{table}. The cut-off needed to be 
decreased somewhat as the couplings were increased
in order to retain numerical stability. As in previous work
(cf. Refs. \cite{WC7,polaron_aniso}), we have checked the accuracy of
our results by comparing the direct evaluation of the kinetic terms
$\Omega_i$ with those obtained with the help of a virial theorem
which relates the $\Omega_i$ to
the potential $V$ through the use of the variational equations:
\be
\Omega_i \Bigr |_{\rm var} \E (-)^i \, \int_0^{\infty} d\sigma \> 
\left [ \, \mu_i^2(\sigma) - \sigma \dot \mu_i^2(\sigma) \, \right ] \, 
\frac{\delta V}{\delta \mu_i^2(\sigma)} \> .
\ee
A relative accuracy of order $10^{-5}$ was achieved for not too large
cut-offs.

\begin{figure}[ht]
\begin{center}
\mbox{\epsfxsize=12cm\epsffile{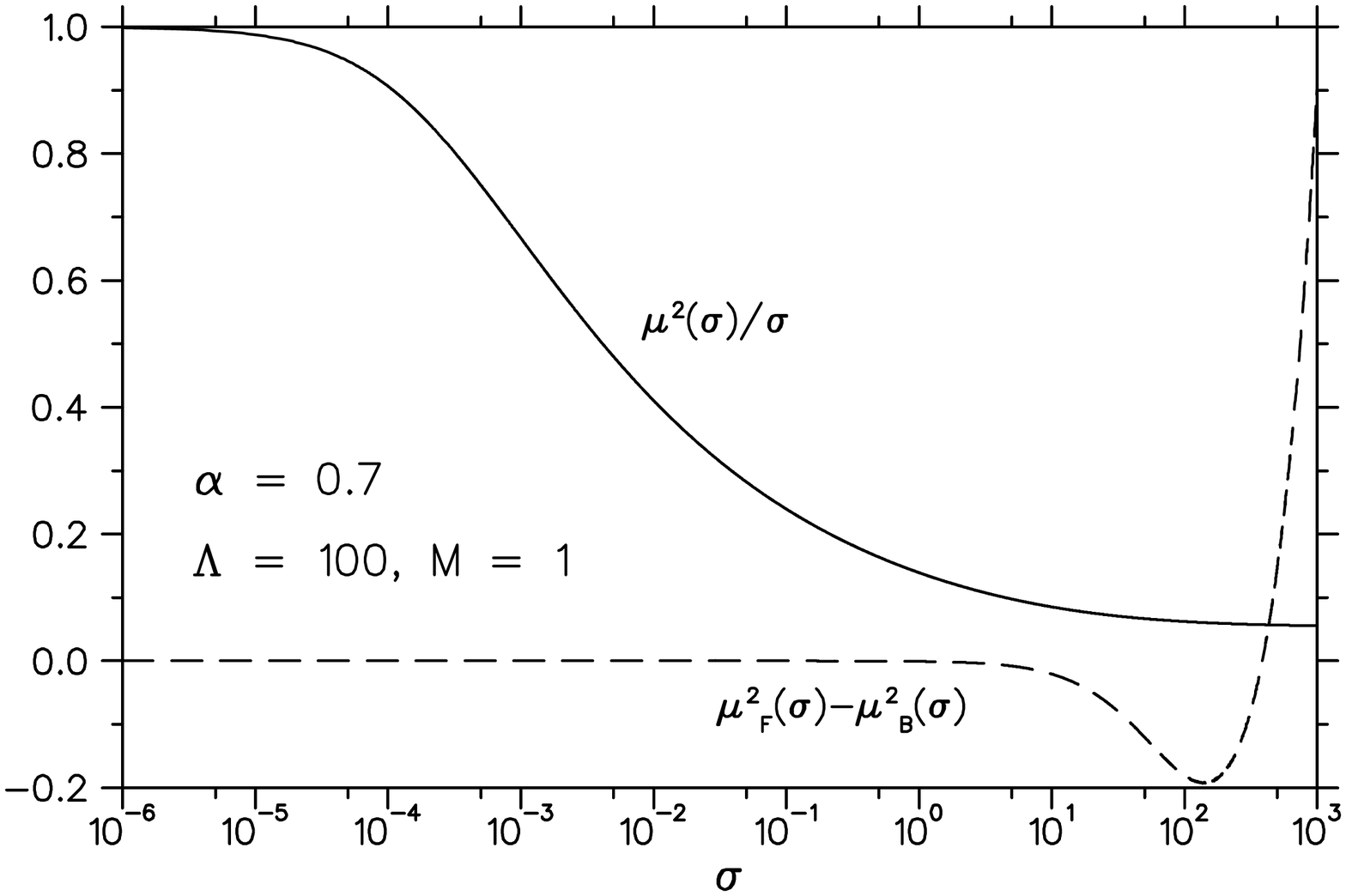}}
\end{center}
\vspace{-1.5cm}
\caption{The pseudotime divided by $\sigma$ (solid line) as
a function of $\sigma$ (on a logarithmic scale) for coupling constant 
$\alpha = 0.7$ and cut-off $\Lambda = 100 \, M$. Bosonic and fermionic 
pseudotime are indistinguishable in this plot. The dashed curve shows 
the difference of both.
}
\label{fig: susy}
\vspace{0.3cm}
\end{figure}

The results in the Table clearly demonstrate, as anticipated in
Sec.~\ref{subsection: kin pot}, that supersymmetry is almost 
perfect.  This may also be seen in
Fig. \ref{fig: susy} where the pseudotimes have been plotted for
one value of $\alpha$ and $\Lambda$.  Differences between the bosonic and 
fermionic pseudotimes only show up in the infrared region
where the less-singular potential contribution $V_2$ 
dominates. This also leads to slightly different values of the bosonic and
fermionic profile function at $ E = 0 $. Our previous conjecture that
SUSY violations would vanish with $\Lambda \to \infty$ \cite{SRA workshop}
turns out to be unfounded: in Table 1 we also give the quantity
\be
\Delta_S \E  - 2 \, \frac{\Omega_B - \Omega_F + V_1}{M_0^2} 
\label{DeltaS}
\ee
which may be considered as a measure
of the importance of SUSY violations in $M_0^2$. One sees that for the 
whole range of coupling constants which we consider they remain
at the percent-level or below

\begin{table}[htb]
\begin{center}
\begin{tabular}{|r|c|c|r|r|c|r||l|l|l|} \hline 
     &     &        &       &       &    &     &         &         &  \\
 $\alpha$  & $\Lambda/M $ & $\lambda$ & $A_B(0)$~~& $A_F(0)$~~& $M_0/M $ 
 & $10^3 \Delta_S$ & \quad $\frac{M_0^2}{M^2 \lambda}$ & $ \> \> \beta(\alpha)$  
& $ \> \lambda A_B(0)$   \\ \hline
     &    &        &       &       &    &   &        &       &   \\
 0.1 &  50& 0.69862& 1.4009& 1.3992& 0.81663& -0.36  &0.95457& 0.95450& 0.97869 \\
     & 100& 0.65402& 1.4964& 1.4946& 0.79011& -0.37  &0.95452&        & 0.97866 \\
     & 300& 0.58901& 1.6615& 1.6595& 0.74981& -0.37  &0.95451&        & 0.97865 \\
     & 500& 0.56102& 1.7444& 1.7423& 0.73178& -0.37  &0.95450&        & 0.97865 \\
     &1000$^{\star}$       & 0.52515& 1.8636& 1.8614 & 0.70800& -0.34 & 0.9545 &
             &0.97865       \\

 0.2 &  50& 0.48548& 1.9802& 1.9716& 0.66592& -1.31  &0.91342& 0.91332& 0.96132 \\
     & 100& 0.42570& 2.2582& 2.2484& 0.62355& -1.38  &0.91335&        & 0.96129 \\
     & 300& 0.34559& 2.7815& 2.7696& 0.56182& -1.38  &0.91333&        & 0.96128 \\
     & 500$^{\star}$       &0.31366& 3.0647 & 3.0516 & 0.53522& -1.42  & 0.9133 & 
             &0.96128      \\

 0.3 &  50& 0.33610& 2.8171& 2.7923& 0.54267& -2.85  &0.87619& 0.87610& 0.94682\\
     & 100& 0.27631& 3.4266& 3.3966& 0.49201& -2.88  &0.87612&        & 0.94681 \\
     & 300& 0.20252& 4.6752& 4.6342& 0.42122& -2.88  &0.87610&        & 0.94680 \\
     & 500$^{\star}$       &0.17527& 5.402~~& 5.355~~&0.39184&  -2.99 & 0.8760 & 
             &0.94680      \\

 0.4 &  50& 0.23214& 4.0256& 3.9682& 0.44224& -4.68  &0.84249& 0.84241& 0.93449 \\
     & 100& 0.17914& 5.2165& 5.1423& 0.38847& -4.71  &0.84242&         & 0.93448 \\
     & 300$^{\star}$       &0.11877& 7.868~~& 7.756~~&0.31629& -4.78   &0.8423  & 
             &0.93448       \\
     & 500$^{\star}$       &0.09810& 9.525~~& 9.390~~&0.28746& -4.81   & 0.8423 & 
             &0.93448      \\
    
 0.5 &  50& 0.16015& 5.769~~& 5.651~~& 0.36059& -6.75 &0.81191& 0.81186& 0.92384 \\
     & 100& 0.11616& 7.953~~& 7.791~~& 0.30709& -6.77  &0.81187&      & 0.92383 \\
     & 300$^{\star}$     & 0.06981 &13.23~~~&12.96~~~ & 0.23805& -6.88 & 0.8118 & 
             &  0.92383   \\

 0.6 &  50& 0.11046& 8.828~~& 8.056~~& 0.29430& -8.95  &0.78410& 0.78406& 0.91451 \\
     & 100& 0.07541&12.127~~ &11.800~~ & 0.24315& -8.98  &0.78406&    & 0.91450 \\
     & 300$^{\star}$     & 0.04117 &22.21~~~ &21.61~~~ & 0.17966& -9.06  & 0.7840 &
             & 0.91450    \\

 0.7 &  50$^{\star}$     & 0.07623 &11.890~~ &11.485~~& 0.24048 & -11.2~~ &0.75867& 
     0.75869 & 0.90626    \\
     & 100$^{\star}$     & 0.04905 &18.48~~~ &17.85~~~ & 0.19290 & -11.2~~ &0.75867& 
             & 0.90625    \\

 0.8 &  50$^{\star}$     & 0.05266 &17.07~~~ &16.37~~~ & 0.19680 & -13.6~~ &0.73540&
     0.73544 & 0.89889    \\
     & 100$^{\star}$     & 0.03198 &28.10~~~ &26.95~~~ & 0.15336 & -13.6~~ &0.73535& 
             & 0.89887    \\

 0.9 &  50$^{\star}$     & 0.03644 &24.49~~~ &23.30~~~ & 0.16130 & -15.9~~ &0.7140 &
     0.71407 & 0.89227    \\
     & 100$^{\star}$     & 0.02091 &42.66~~~ &40.60~~~ & 0.12219 & -15.9~~ &0.7139 & 
             &  0.89225  \\

 1.0 &  50$^{\star}$     & 0.02526 &35.08~~~ &33.13~~~ & 0.13244 & -18.2~~ &0.6943~& 
     0.69435 &  0.88627  \\
     & 100$^{\star}$     & 0.01372 &64.61~~~ &61.01~~~ & 0.09757 & -18.3~~ &0.6941~& 
             &  0.88622  \\
     &     &        &       &       &   &     &        &        &  \\ \hline
\end{tabular}

\end{center}
\caption{Some results of the numerical solution of the variational 
equations with form factor regularization for different cut-offs and 
coupling constants. A relative accuracy of $10^{-6}$ was required at 
each point between subsequent iterations; in entries marked by a 
$^{\star}$ only $10^{-5}$ relative accuracy was achieved. 
The quantity $\Delta_S$ measures the SUSY violation and 
is defined in Eq. (\ref{DeltaS}).
The last three columns compare the numerical 
results with analytical approximations explained in the text.}
\label{table}
\end{table}
\clearpage

\unitlength1mm

\noindent
and become cut-off independent within 
numerical accuracy.

\begin{figure}[ht]
\begin{center}
\mbox{\epsfxsize=12cm\epsffile{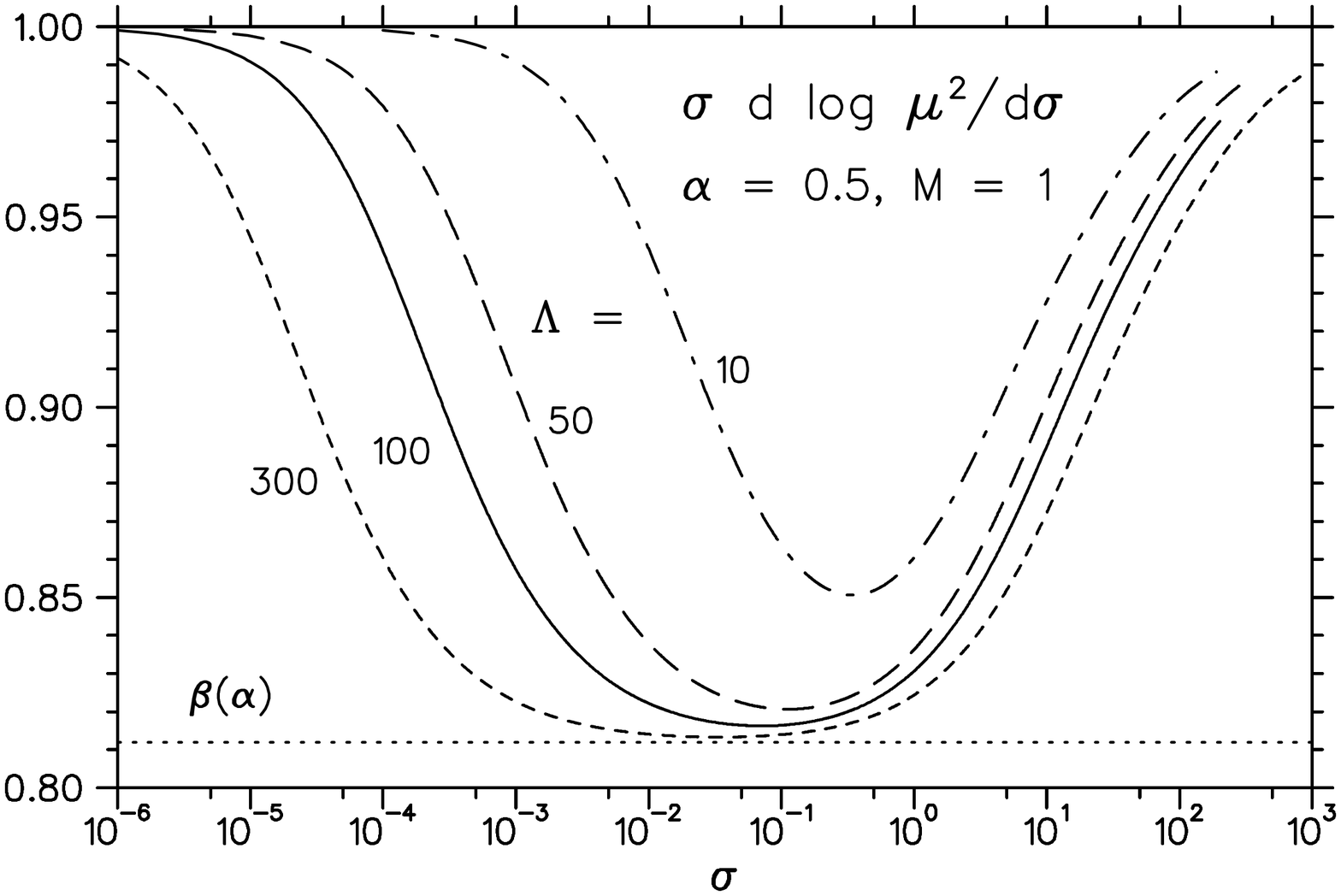}}
\end{center}
\vspace{-1.8cm}
\caption{The effective power of $\sigma$ for the the bosonic 
pseudotime $\mu^2(\sigma)$ from 
numerical solutions of the variational Abraham-Lorentz equation for QED. 
As in Fig.~\ref{fig: susy}, the difference between $\mu^2_F(\sigma)$ and 
$\mu^2_B(\sigma)$ is too small to show up on this plot.  Results for 
various cut-offs $\Lambda$ in units of the electron mass $M$ are shown. 
The dashed horizontal line denotes the numerical value from the 
analytical solution \protect{(\ref{transc eq for beta})} at coupling
constant $\alpha = 0.5$.
}
\label{fig: power}
\vspace{0.3cm}
\end{figure}

The ratio $ \> \sigma \dot \mu^2(\sigma)/\mu^2(\sigma)$ is plotted in
Fig. \ref{fig: power} for $\alpha = 0.5$ and various
cut-offs. Independently of the actual magnitude of $\mu^2(\sigma)$,
power-like behaviour of the pseudotime corresponds to a horizontal
line on this plot.  Because $\mu^2(\sigma) \stackrel{\sigma\to
0}{\longrightarrow} \sigma$ and $\mu^2(\sigma)
\stackrel{\sigma\to\infty}{\longrightarrow} \sigma/A(0)$ all curves go
towards unity for small and large $\sigma$.  What is interesting,
however, is that for an increasing range of intermediate
$\sigma$-values (as $\Lambda \to \infty$) the pseudotimes exhibit
power-like behaviour of the form $\sigma^{\beta},\>\beta < 1$.
Moreover, the value of $\beta$ seems to become independent of the
cut-off as $\Lambda$ increases.

\unitlength1mm

This result can be understood analytically: for large cut-offs and 
moderate values of $\sigma$ the approximate VALEs (\ref{AL 4}) 
and~(\ref{AL 5}) should become valid.  Inserting the {\it ansatz} 
\be
\tilde \mu^2(\sigma) \To  s_0 \cdot \sigma^{\beta} 
\label{power ansatz}
\ee
into Eq. (\ref{AL 5}) one obtains
\be
s_0 \, \sigma^{\beta} + \frac{3 \alpha}{4 \pi} \, \int_0^{\infty} 
d\sigma' \> \left [
\, \left ( 1 + \frac{\sigma}{\sigma'} \right )^{\beta} + 
\left | 1 - \frac{\sigma}{\sigma'} \right |^{\beta} - 2 \, \right ] \E 
\sigma + \frac{\kappa_E}{\Lambda^2}\> .
\ee
Scaling $\sigma' = t \sigma$ shows that the integral is proportional
to $\sigma$ and therefore dominates the l.h.s. for large values of
$\sigma$ provided $\beta < 1 $ which turns out to be the case for 
{\it positive} couplings \footnote{For $\alpha < 0 $ no power-like 
solutions exist anymore as the term $\sigma^{\beta}$ would dominate 
for $\beta > 1 $.}.
Hence, for large $\sigma$, the power $\beta$ is only a
function of the coupling constant $\alpha$ and determined by the
implicit equation
\be
\int_0^{\infty} dt \> \left [
\, \left ( 1 + \frac{1}{t} \right )^{\beta} + 
\left | 1 - \frac{1}{t} \right |^{\beta} - 2 \, \right ] \E 
\frac{4 \pi}{3 \alpha} \> .
\ee
After performing the integral one obtains the following
transcendental equation for $\beta(\alpha)$
\be
\frac{\pi}{2} \beta \, \tan \left (\frac{\pi}{2} \beta \right ) \E 
\frac{2 \pi}{3 \alpha} \> .
\label{transc eq for beta}
\ee
Numerical solutions for a variety of $\alpha$'s have been tabulated in
Table~\ref{table}.
For small coupling constants the solution to this equation behaves like
\be
\beta \to 1 - \frac{3 \alpha}{2 \pi} + \left ( \frac{3 \alpha}{2 \pi} 
\right )^2 + \ldots 
\label{beta pert}
\ee
confirming that for positive coupling constant $\beta < 1 $, while for 
large $\alpha$ it goes like $ \beta \to \sqrt{8/(3 \pi \alpha)} \to 0 $.  
Fig. \ref{fig: power} shows that the value $\beta(0.5) = 0.81186$ 
(see  Table~\ref{table}) explains the intermediate-range behaviour of 
the pseudotime very well. For very large $\sigma$ the ASUSY 
approximation~(\ref{AL 5}) becomes invalid and the 
pseudotime reverts to linear behaviour.

\section{The anomalous mass dimension}

The power-like behaviour of $\mu^2(\sigma)$ is also the key for 
obtaining the variational approximation for the anomalous mass dimension 
regularized via a form factor rather than dimensionally~\cite{QED2}.
To do this we first note that in $V$ the variational parameter
$\lambda$ occurs exclusively in the combination $\lambda M$.  Let us,
therefore, define the alternative dimensionless variational parameter 
$x_{\lambda} = \lambda M/\Lambda$ so that Mano's equation becomes
\be
M_0^2 \E \Lambda^2 \left [ 2 x_\lambda \frac{M}{\Lambda} \> - \>
 x_\lambda^2 \> - \> 2\, H(\alpha,x_\lambda)\right ]\;\;\;,
\label{eq: new mano}
\ee
where $H(\alpha,x_\lambda)$ is the dimensionless combination 
$(\Omega_B-\Omega_F+V)/\Lambda^2$.  Note that, by construction,
$H(\alpha,x_\lambda)$ no longer depends on $M$ explicitly.  Therefore,
as it is dimensionless, it can also no longer depend on $\Lambda$
explicitly.  The dependence on the cut-off can only enter implicitly
through the variational parameters and functions.   However, the r.h.s.
of Eq.~(\ref{eq: new mano}) is stationary with respect to variational 
parameters/functions, so that the dependence of $M_0^2$ on $\Lambda$, at
fixed $M$, is just given by the explicit $\Lambda$ dependence in 
Eq.~(\ref{eq: new mano}), i.e.
\be
\frac{\partial M_0^2}{\partial \Lambda} \Biggr |_{M \> {\rm fixed}} \E 
2 \frac{M_0^2}{\Lambda} - 2 M \, x_{\lambda} \> .
\label{flow eq}
\ee
Writing $M_0 =  Z_M \, M_{\nu} $ (where $\nu$ is an
arbitrary mass parameter), the anomalous mass dimension
may now be evaluated with the help of this flow equation as
\be
\gamma_M \E  
\frac{\partial \log Z_M}{\partial \log \nu} \E 
 - \frac{\partial \log M_0}{\partial \log \Lambda} \E 
 - \frac{\Lambda}{2 M_0^2} \left ( \, 2 \frac{M_0^2}{\Lambda} - 2 M \, 
x_{\lambda} \, \right ) \E \frac{M^2}{M_0^2} \lambda - 1 \> .
\label{gamma M from flow eq}
\ee
Of course, the (existence of the) limit $\Lambda \to \infty$ is 
understood in the above equations. If this limit exists then $\gamma_M$ 
must  
necessarily be cut-off independent so that a further differentiation of 
Eq. (\ref{gamma M from flow eq}) with respect to $\Lambda$ gives
\be
0 \E - \frac{M^2}{M_0^4} \frac{\partial M_0^2}{\partial \Lambda} \lambda 
+ \frac{M^2}{M_0^2} \, \frac{\partial \lambda}{\partial \Lambda} \> .
\ee
Using Eqs. (\ref{flow eq}) and (\ref{gamma M from flow eq}) we obtain 
$ \, \Lambda \, \partial \lambda/\partial \Lambda = - 2 \gamma_M \, 
\lambda \, $ and integration then shows how the parameter $\lambda$ 
behaves for very large cut-offs
\be
\lambda \> \stackrel{\Lambda \to \infty}{\longrightarrow} \> 
{\rm const.} \, \left ( \frac{M^2}{\Lambda^2} \right )^{\gamma_M} \> .
\label{lambda(Lambda)}
\ee
This behaviour comes as no surprise  since we know~\cite{WC7} that in the 
worldline formalism the
bare and effective mass of the quantum mechanical particle are 
$\kappa_E$ and $\kappa_E/\lambda$, respectively,  for which a similar 
relation as for the actual masses is expected.
Thus the anomalous mass dimension can be determined either numerically by
solving the variational equations and evaluating 
Eq.~(\ref{gamma M from flow eq}) for larger and larger cut-offs or 
analytically from Eq.~(\ref{lambda(Lambda)}) by finding the cut-off 
dependence of $\lambda$. 

In the following we will pursue the latter option which is possible as we
know the approximate behaviour of the pseudotime for small, 
intermediate and large values of $\sigma$ (here we take for simplicity 
$ \kappa_E = 1$)
\be
\tilde \mu^2(\sigma) \> \simeq \> \left ( \sigma + \frac{1}{\Lambda^2} 
\right ) \, \Theta \left ( \sigma_1 -\sigma \right ) + 
s_0 \sigma^{\beta} \, \Theta \left ( \sigma - \sigma_1\right)  
\Theta \left ( \sigma_2  - \sigma \right ) 
+ \frac{\sigma}{A(0)} \, \Theta \left ( \sigma  - \sigma_2 \right ) \> .
\ee
The regions are separated by $\sigma_1 = x_1/\Lambda^2 $, where the 
cut-off $1/\Lambda^2$ becomes effective in $\tilde \mu^2$,
and $\sigma_2 = 2 x_2/(\lambda^2 M^2 A(0)) $ where the exponential 
$\exp(-\tilde \gamma)$ becomes important and the ASUSY approximation 
breaks down. $x_1, x_2$ are numbers of order one which can be 
determined (together with the 
normalization factors $A(0)$ and $s_0$) by matching the approximate 
solutions at the boundaries. As we only need the leading terms in the 
cut-off $\Lambda$  in order to determine $\gamma_M$, the actual numerical 
values of $x_1, x_2$ do not matter. 
Matching gives $s_0 \sim (\Lambda^2)^{\beta - 1}$ 
as expected from dimensional arguments and
\be
A(0) \> \sim \> \left ( \frac{\Lambda^2}{\lambda^2 M^2} 
\right )^{\frac{1-\beta}{2-\beta}} \> .
\label{A0 approx}
\ee
We now insert these expressions into the variational equation 
for $\lambda$ in SUSY approximation
\be
\frac{1}{\lambda} \E 1 + \frac{3 \alpha}{\pi} \, 
\int_0^{\infty} d\sigma \>\frac{1}{\tilde \mu^2(\sigma)}
 \>
\frac{(1+\tilde \gamma+\tilde \gamma^2)e^{-\tilde \gamma} -1}
{\tilde \gamma^2}
 \;\;\;.
\label{var eq lambda 2}
\ee
As $\tilde \gamma$ only becomes large in the region $\sigma > \sigma_2$
we may expand the exponential in the intervals for small and medium values
of $\sigma$. After performing the 
integrations, one obtains
\be
\frac{1}{\lambda} \E 1 + \frac{3 \alpha}{\pi}
\left\{\left[ \frac{1}{2}\ln(1+x_1)\right] \> + \> 
 \frac{1}{2 (1-\beta)}\left[ A(0) - \frac{x_1}{1+x_1}\right]
\> + \> \left[ A(0)\,\cdot \,{\rm const}\right ]\right \}\;\;\;.
\label{eq: approx lam eq}
\ee
If $A(0)$ would decrease with increasing $\Lambda$ (or stay constant)
this equation would imply that $\lambda$ would go to a constant.  This, 
however, is inconsistent with Eq.~(\ref{A0 approx}).  Hence  $A(0)$ must
increase with $\Lambda$, with Eq.~(\ref{eq: approx lam eq}) implying that
$\lambda A(0)$ remains constant.  This behaviour is clearly seen in the
numerical results shown in the last column of Table~\ref{table}.  Also,
in this case the low-$\sigma$ region can be neglected, which is
reasonable because otherwise the final result would depend on the 
details of the regularization.
Comparing this with Eqs.~(\ref{lambda(Lambda)}) and~(\ref{A0 approx}) 
one sees that consistency requires that
\be
\gamma_M = \frac{1-\beta}{\beta}\;\;\;
\label{eq: bet gam rel}
\ee
and so Eq.~(\ref{gamma M from flow eq}) implies that
\be
\frac{M_0^2}{M^2\, \lambda} \E\beta\;\;\;.
\ee
Comparison of the $8^{\rm th}$ and $9^{\rm th}$ columns in 
Table~\ref{table} shows clear numerical evidence that this relation is 
indeed fulfilled. Note also that Eqs.~(\ref{transc eq for beta}) 
and~(\ref{eq: bet gam rel}) imply 
that $\gamma_M$ is a solution of the implicit equation
\be
\frac{3}{4} \alpha \E \left ( 1 + \gamma_M \right ) \, \tan \left ( 
\frac{\pi}{2}\,\frac{\gamma_M }{1 + \gamma_M} \right ) \> ,
\label{transc eq for gamma}
\ee
in agreement with the result obtained in dimensional regularization
\cite{QED2}.  The convergence of the perturbative expansion and the
analytic properties of the solutions of such transcendental equations
have been studied in Ref. \cite{MRS}. The present derivation
adds the insight that for $\alpha < 0 $ no power-like solution of the 
VALE (\ref{AL 5}) exists and therefore Eq. (\ref{transc eq for beta}) 
is not applicable anymore. Obviously, this peculiar behaviour when 
changing the sign of the coupling constant is {\it not} contained in 
Eq. (\ref{transc eq for gamma}) but in agreement with Dyson's old 
qualitative argument \cite{Dys} that $\alpha = 0$ is an essential 
singularity in QED. We also note that Dyson-Schwinger calculations
with a particular {\it ansatz} for the electron-photon vertex lead to
a similar implicit equation \cite{DS2} but a gauge dependence remains
and solutions exist only below a critical coupling.

\vspace{1cm} 
\section{Conclusions and outlook}
In summary, we have shown that the variational formulation of worldline
QED very naturally leads to an equation which is similar to the one
considered much earlier by Abraham, Lorentz and Dirac in attempts to
describe the electron and its self-interaction with the radiation field.
In contrast to these attempts our approach contains (almost) all the
ingredients of the relativistic field theory of electrons and photons, 
in particular its divergence structure. This has been demonstrated by 
numerically solving the variational Abraham-Lorentz equation (VALE) for 
a variety of cut-offs and by deriving an approximate nonperturbative 
expression for the anomalous mass dimension of the electron. We have 
shown how the approach leads naturally to qualitatively different 
behaviour of the theory for $\alpha>0$ and $\alpha<0$.  Furthermore,
while the present investigation has been restricted to a free
electron interacting with its own radiation field, the extension to the 
case when an external field is present as well should be 
straight-forward. This would allow a study of how this 
field-theoretically based worldline variational approach to QED avoids 
the pitfalls of pre-acceleration and acausality which have plagued all 
classical attempts.

However, for further progress we deem it more important
to first study the issue of supersymmetry (breaking) within the 
worldline variational approach in more detail.
For massive electrons small violations of worldline supersymmetry
have been observed which are ``soft'' in the sense that the ultraviolet 
behaviour of the theory is not affected. These violations presumably 
reflect the SUSY-breaking generated by the different boundary conditions 
for bosonic and fermionic variables in the exact action.
Further investigation is required into what role, if any, this 
SUSY-breaking plays in the full spin structure of the electron 
propagator and how it is manifested in the worldline variational 
approximation. Such an understanding is required for future applications 
of this non-perturbative approach to physical processes.
Finally it is not inconceivable that similar VALE's
as derived here in the variational approach for the propagator
will also emerge for the full interacting vertex.

\vspace{5cm} 
\bce
{\Large\bf Appendix}
\ece

\renewcommand{\thesection}{\Alph{section}}
\renewcommand{\theequation}{\thesection.\arabic{equation}}

\setcounter{section}{0}

\section{Approximate  $3^{\rm rd}$-order differential VALE}
\label{app: 3rd order de}
\setcounter{equation}{0}

Here we show how a $3^{\rm rd}$-order  
differential equation for the mean square displacement arises 
approximately in QED. 
For this purpose we invert Eq. (\ref{def pseudotime}) to obtain
\be
\frac{1}{A_i(E)} \E 1 + \int_0^{\infty} d\sigma \, 
\ddot \mu^2_i(\sigma) \, \cos (E \sigma) 
\label{A from mu2}
\ee
and observe that {\it for small coupling} the profile function stays 
close to unity. Therefore one has in this case
\bea
\Omega_i \EA  \frac{2 \kappa_E}{\pi} \int_0^{\infty} dE \, \left [ \, 
- \log \left ( 1 + \frac{1}{A_i(E)} - 1 \right ) 
+ \frac{1}{A_i(E)} - 1  \right ] \non
\EA \frac{\kappa_E}{\pi} \int_0^{\infty} dE \, \left \{ \, 
\left [ 1 - \frac{1}{A_i(E)} \right ]^2 + {\cal O} \left ( 1 - 
\frac{1}{A_i(E)} 
\right )^3 \, \right \} \simeq \frac{\kappa_E}{2} \int_0^{\infty} 
d\sigma \, \left [ \, \ddot \mu^2_i(\sigma) \, \right ]^2 \> .
\label{Omega approx}
\eea
Variation first gives a fourth-order ordinary differential 
equation 
\be
(-)^i \, \kappa_E \, \frac{d^4}{d \sigma^4} \, \mu^2_i (\sigma) + 
\frac{\delta V}{\delta \mu_i^2 (\sigma)} \> \simeq \> 0 \> ,
\label{4th order approx}
\ee
but in the supersymmetric limit of QED the functional derivative of 
the ``potential'' $V$ is a total derivative (see 
Eq. (\ref{var eq for muF})) so that an integration yields
\be
\stackrel{\ldots}{\mu}^2(\sigma) \> \simeq \> \frac{3 \alpha}{2 \pi} \, 
\frac{\dot \mu^2(\sigma)}{\tilde \mu^4(\sigma)} \, 
e^{- \tilde \gamma(\sigma)} \> .
\label{3rd order approx 1}
\ee
This $3^{\rm rd}$ order differential equation clearly is specific for 
QED and therefore it is tempting to interpret the ``kinetic'' term 
$\Omega$ in this case as total (integrated) ``radiative energy loss''. 
Indeed, it is positive and in the approximate form  (\ref{Omega approx}) 
the integrand has precisely the form of the
Larmor power formula (see, e.g. Eq. (17.6) in Ref. \cite{Jack}).
Without external force there should be, of course, no real 
radiation loss -- in fact, for exact supersymmetry $\Omega_B$ is 
completely cancelled by $\Omega_F$. Of course, the $4^{\rm th}$ order 
equation also applies (approximately) 
to other theories where this interpretation does not make sense.

\vspace{0.2cm}
Since we have assumed $\ddot \mu^2$ as small in Eq. (\ref{A from mu2}) 
the present approximation should be similar to a derivative expansion. 
Indeed, if in the integrand of the VALE (\ref{AL 4})
\be 
\mu^2 \left (\sigma + \sigma' \right ) - \mu^2 \left (|\sigma - \sigma'| 
\right) \E \mu^2 \left (\sigma_> + \sigma_< \right ) - 
\mu^2 \left (\sigma_> - \sigma_< \right) \> \simeq \> 
\dot \mu^2 \left ( \sigma_> \right ) \, \sigma_<
\ee
is expanded to first order and the 
resulting equation divided by $\sigma$ one obtains 
\be
\frac{\dot \mu^2(\sigma)}{\sigma} \, \left [ \, 1 + \frac{3 \alpha}{2 \pi}
\int_0^{\sigma} d\sigma' \, \sigma' 
\frac{\dot \mu^2(\sigma')}{\tilde \mu^4(\sigma')} 
\, e^{-\tilde \gamma(\sigma')} \, \right ] 
+ \frac{3 \alpha}{2 \pi} \,
\int_{\sigma}^{\infty} d\sigma' \, \frac{\dot \mu^4(\sigma')}{\tilde 
\mu^4(\sigma')} \, e^{-\tilde \gamma(\sigma')}
\> \simeq \>  \frac{1}{\sigma} \> .
\ee
This can be converted to a differential equation by differentiating w.r.t.
$ \> \sigma$ and replacing the square bracket 
by $1/(\dot \mu^2 - \sigma \ddot \mu^2) $.
Another differentiation then gives
\be
\stackrel{\ldots}{\mu}^2(\sigma) \> \simeq \> \frac{3 \alpha}{2 \pi} \, 
\left [ \, \dot \mu^2(\sigma)  - \sigma \ddot \mu^2(\sigma) \, 
\right ]^2 \, \frac{\dot \mu^2(\sigma)}{\tilde \mu^4(\sigma)} \, 
e^{- \tilde \gamma(\sigma)} 
\label{3rd order approx 2}
\ee
which is identical with the third-order equation 
(\ref{3rd order approx 1}) if the factor in square brackets is replaced 
by its perturbative value $1$.
The boundary conditions are $ \mu^2(0) = 0 \, , \, \dot \mu^2(0) = 1 $ 
and a peculiar value of the second derivative at the origin
\be
\ddot  \mu^2(0) \E   \frac{1}{\kappa_E} \, \int_0^{\infty} d\sigma \> 
\mu^2_F(\sigma) \, \frac{\delta V}{\delta \mu^2_F(\sigma)} 
\Biggr |_{\rm SUSY} \E - \frac{3 \alpha}{2 \pi} \, \int_0^{\infty} 
d\sigma \> \left ( \frac{\dot \mu^2(\sigma) }{\tilde \mu^2(\sigma)} 
\right )^2 \, e^{-\tilde \gamma(\sigma)}
\ee
expressed as functional of the still unknown solution. 
This follows from Eq. (\ref{AL 3}) in the limit $\sigma \to 0$, 
together with Eq. (\ref{var eq for muF}) and can be considered
as a reminder that the nonlocal actions we are working with originate 
from local actions with the photonic degrees of freedom integrated out. 
In this way Ostrogradski's no-go theorem for higher derivative and 
nonlocal theories \cite{BNW} is evaded.

It is interesting that the (approximate) equation 
(\ref{3rd order approx 2}) also allows for power-like solutions in the 
massless case whereas the perturbative version 
(\ref{3rd order approx 1}) does not.
Indeed, inserting the {\it ansatz} (\ref{power ansatz}) into Eq. 
(\ref{3rd order approx 2}) shows that the powers on both sides 
of the equation match and leads to the cubic equation
\be
\beta - 1  \> \simeq \> \frac{3 \alpha}{2 \pi} \, \beta^2 (\beta - 2)
\label{cubic eq for beta} \> .
\ee
This gives the correct limit (\ref{beta pert}) for $\alpha \to 0$
and the same $ 1/\sqrt{\alpha}$-behaviour for large $\alpha$ as the exact
result (\ref{transc eq for beta}) only with a coefficient
$\pi/\sqrt{8} = 1.11$ times larger. 
That the derivative expansion works rather well also at intermediate 
coupling constants is demonstrated, for example, by the numerical value 
$\beta(0.5) \, \simeq \, 0.81277$ 
obtained from Eq. (\ref{cubic eq for beta}) to be compared with 
the correct $\beta(0.5) = 0.81186$.

\end{document}